\documentclass{article}
\usepackage{dcase2022, amsmath, graphicx, url, times, booktabs, tabularx}
\usepackage{multirow}
\usepackage{textcomp}
\usepackage{amsmath}
\usepackage{graphicx}
\usepackage{color}
\usepackage{booktabs}
\usepackage{colortbl}

\title{Language-based Audio Retrieval Task in DCASE 2022 Challenge}

\name{Huang Xie, Samuel Lipping, Tuomas Virtanen}
\address{Audio Research Group, Tampere University, Tampere, Finland}

\begin{document}

    \ninept
    \maketitle

    \begin{sloppy}

        \begin{abstract}
            Language-based audio retrieval is a task, where natural language textual captions are used as queries to retrieve audio signals from a dataset.
            It has been first introduced into DCASE 2022 Challenge as Subtask 6B of task 6, which aims at developing computational systems to model relationships between audio signals and free-form textual descriptions.
            Compared with audio captioning (Subtask 6A), which is about generating audio captions for audio signals, language-based audio retrieval (Subtask 6B) focuses on ranking audio signals according to their relevance to natural language textual captions.
            In DCASE 2022 Challenge, the provided baseline system for Subtask 6B was significantly outperformed, with top performance being 0.276 in mAP@10.
            This paper presents the outcome of Subtask 6B in terms of submitted systems\textquotesingle~performance and analysis.
        \end{abstract}

        \begin{keywords}
            Language-based audio retrieval, DCASE 2022 Challenge, Clotho.
        \end{keywords}

        \section{Introduction}\label{sec:intro}

        With the growth of multimedia content in recent decades, there is a need for retrieval methods that can efficiently organize the data based on its content, and retrieve relevant items when doing searches to datasets.
        Natural language provides an efficient way to represent complex information about multimedia.
        It can represent high level information about data that goes beyond any fixed taxonomies.
        For audio signals, natural language can represent information related to temporal relationships between sound sources, and attributes of sounds and their environment.

        Language-based multimedia retrieval has received increasing attention in recent years.
        The majority of recent works has focused heavily on the visual domain~\cite{Dong2021Dual, Qu2021Dynamic}.
        For example, there are plenty of approaches~\cite{Kaur2021Comparative} tackling content-based image retrieval with free-form textual descriptions.
        In contrast, only a few studies have been conducted on language-based audio retrieval in the existing literature.
        Early works~\cite{Chechik2008Large, Elizalde2019Cross} deal with language-based audio retrieval using multi-word text queries consisting of audio tags or class labels, rather than sentence-like textual descriptions, e.g., captions.
        Recent studies~\cite{Oncescu2021Audio, Xie2022Unsupervised} prompt research in this field by exploring human written captions as queries.
        In DCASE 2022 Challenge, language-based audio retrieval is introduced into as Subtask 6B, which aims to inspire further research into audio retrieval with unconstrained textual descriptions.

        In this paper, we present the task setup and submissions for Subtask 6B of task 6 in DCASE 2022 Challenge.
        We introduce the datasets for system development and evaluation, describe the baseline system for Subtask 6B, and present the challenge submissions.
        Evaluation and analysis of submitted systems includes general statistics on systems and performance and system characteristics.

        The remainder of this paper is organized as follows.
        In Section~\ref{sec:task-setup}\@, we describe the task setup, including task description, task datasets, and evaluation metrics.
        Then, we introduce the task baseline system in Section~\ref{sec:baseline-system}\@.
        We present the evaluation results and analysis of challenge submissions in Section~\ref{sec:challenge-submissions}\@.
        Finally, we conclude this paper in Section~\ref{sec:conclusions}\@.

        \begin{figure*}[!t]
            \centering
            \includegraphics[width=0.9\textwidth]{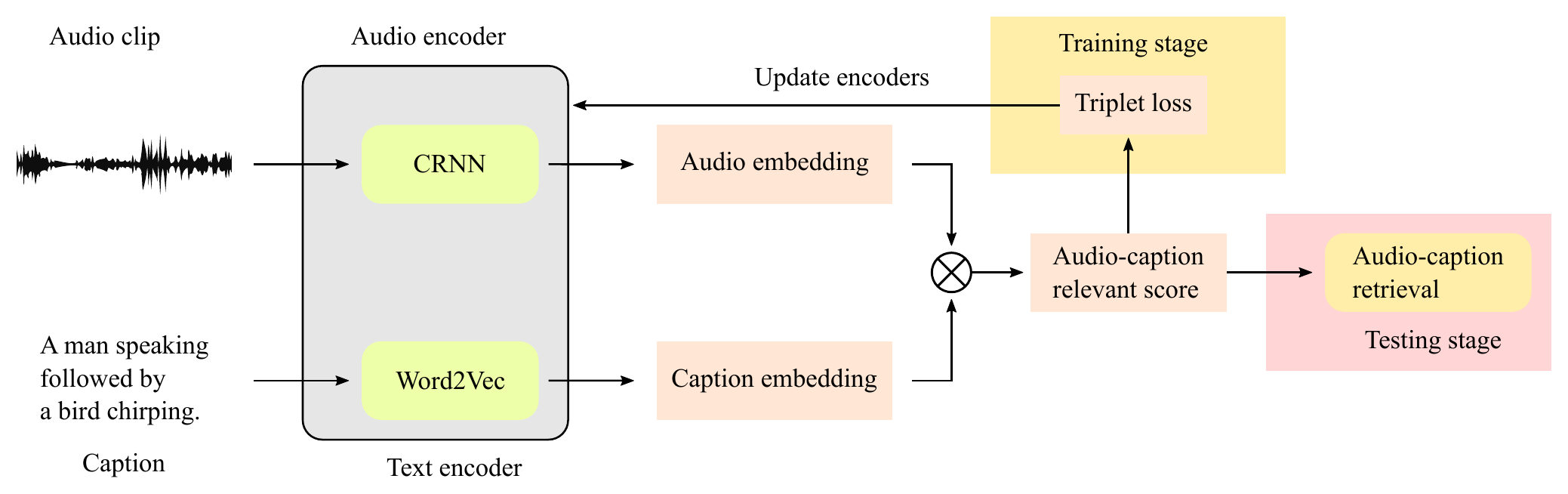}
            \caption{The baseline system for the language-based audio retrieval subtask.}
            \label{fig:baseline_system}
        \end{figure*}

        \section{Task Setup}\label{sec:task-setup}

        In this section, we introduce the task description, datasets for system development and evaluation, and evaluation metrics.

        \subsection{Task Description}\label{subsec:task-description}

        Language-based audio retrieval is concerned with retrieving audio signals using their sound content textual descriptions (i.e., audio captions).
        With this task, the goal is to evaluate audio retrieval methods, where a retrieval system takes an audio caption as a text query and ranks audio signals in a fixed dataset according to their relevance to the caption.
        In DCASE 2022 Challenge, human-written natural language audio captions are used as text queries.
        For each query, the retrieval task is to retrieve 10 audio files from a given evaluation dataset and sort them according to their relevance to the query.

        \subsection{Development Dataset}\label{subsec:development-dataset}

        The Clotho v2~\cite{Drossos2020Clotho} is provided as the task development dataset, which consists of audio samples of 15 to 30 seconds duration, with each audio sample having five captions of eight to 20 words length.
        There are 6,974 audio samples with 34,870 captions in total.
        All audio samples are sourced from the Freesound platform~\cite{Font2013Freesound}, and captions are crowd-sourced using a three-step framework~\cite{Drossos2020Clotho}.

        The Clotho v2~\cite{Drossos2020Clotho} is divided into a training split of 3,839 audio clips with 19,195 captions, a validation split of 1,045 audio clips with 5,225 captions, and a testing split of 1,045 audio clips with 5,225 captions.
        These splits are created by first constructing the sets of unique words of the captions of each audio clip.
        These sets of words are combined to form the bag of words of the whole dataset, from which the frequency of a given word can be derived.
        With the unique words of audio files as classes, multi-label stratification is applied.
        The data collecting procedure is explained in detail in~\cite{Drossos2020Clotho}.

        \subsection{Clotho Retrieval Evaluation Dataset}\label{subsec:clotho-retrieval-evaluation-dataset}

        The task evaluation dataset consists of 1,000 audio samples sourced from the Freesound platform~\cite{Font2013Freesound}, and one human written caption is provided for each audio sample.
        The audio samples are collected following the procedure described in~\cite{Drossos2020Clotho}, by optimizing the tag distribution of the selected samples.
        The samples were selected from the set of files not used in Clotho v2.
        Captions were gathered using the first crowd-sourcing step of Clotho v2 manually screened for typographical errors and speech transcription.
        Table~\ref{tab:datasets} summarizes the information about the development and evaluation datasets.

        \begin{table}[!t]
            \centering
            \begin{tabular}{c|c|c|c}
                \hline
                \bfseries Dataset                      & \bfseries Split & \#\bfseries Audio & \#\bfseries Captions \\
                \hline
                \multirow{3}{*}{\bfseries Development} & Training        & 3839              & 19195                \\
                \cline{2-4}
                & Validation      & 1045              & 5225                 \\
                \cline{2-4}
                & Testing         & 1045              & 5225                 \\
                \hline
                \multicolumn{2}{c|}{\bfseries Evaluation} & 1000 & 1000 \\
                \hline
            \end{tabular}
            \caption{Statistics of the development and evaluation datasets.}
            \label{tab:datasets}
        \end{table}

        \subsection{Evaluation Metrics}\label{subsec:evaluation-metrics}

        In the evaluation, the ground truth relevance of audio samples are considered binary (i.e., only the audio samples belong to the caption query are considered relevant, and all the others not relevant).
        The submissions for this task will be evaluated using mean average precision at top-10 (mAP@10) as the main metric, and recall at $k$ (R@\textit{k} with \textit{k} $\in \{1, 5, 10\}$) as the secondary metrics.

        The mAP metric has been widely used for evaluating the performance of cross-modal retrieval algorithms~\cite{Kaur2021Comparative}\@.
        It is a rank-aware metric, which measures the mean of average precisions (AP) over all the queries.
        The AP for a query is calculated by averaging the precisions at positions, where relevant items are in the retrieved rank list.
        The more relevant items in the top rank list, the higher mAP value it has.
        The R@\textit{k} metric is another standard, rank-unaware retrieval metric~\cite{Oncescu2021Audio}, which is defined as the proportion of relevant items among the top-K retrieved results to all the relevant items in the evaluation dataset, averaged across all the caption queries.
        The challenge submissions will be ranked by the mAP@10 metric.

        \begin{table*}[!t]
            \centering
            \begin{tabular}{rr||cc|cc|cc|cc}
                \toprule
                \textbf{Rank} & \textbf{Team} & \multicolumn{2}{c|}{\textbf{mAP@10} with 95\% CI} & \multicolumn{2}{c|}{\textbf{R@1} with 95\% CI} & \multicolumn{2}{c|}{\textbf{R@5} with 95\% CI} & \multicolumn{2}{c}{\textbf{R@10} with 95\% CI} \\
                \midrule
                1  & Xu et al.~\cite{xu2022_t6b}          & 0.276 & [0.254, 0.299] & 0.176 & [0.152, 0.200] & 0.416 & [0.385, 0.447] & 0.536 & [0.505, 0.567]  \\
                2  & Mei et al.~\cite{mei2022_t6b}        & 0.251 & [0.229, 0.273] & 0.153 & [0.131, 0.175] & 0.387 & [0.357, 0.417] & 0.504 & [0.473, 0.535]  \\
                3  & Lamort et al.~\cite{lamort2022_t6b}  & 0.221 & [0.200, 0.242] & 0.131 & [0.110, 0.152] & 0.343 & [0.314, 0.372] & 0.466 & [0.435, 0.497]  \\
                4  & Pellegrini~\cite{pellegrini2022_t6b} & 0.216 & [0.195, 0.237] & 0.127 & [0.106, 0.148] & 0.321 & [0.292, 0.350] & 0.463 & [0.432, 0.494]  \\
                5  & Lai et al.~\cite{lai2022_t6b}        & 0.215 & [0.194, 0.235] & 0.122 & [0.102, 0.142] & 0.328 & [0.299, 0.357] & 0.478 & [0.447, 0.509]  \\
                6  & Wu et al.~\cite{wu2022_t6b}          & 0.188 & [0.168, 0.207] & 0.107 & [0.088, 0.126] & 0.303 & [0.275, 0.331] & 0.413 & [0.382, 0.444]  \\
                7  & Weck et al.~\cite{weck2022_t6b}      & 0.128 & [0.111, 0.145] & 0.077 & [0.060, 0.094] & 0.188 & [0.164, 0.212] & 0.284 & [0.256, 0.312]  \\
                8  & Xiao et al.~\cite{xiao2022_t6b}      & 0.097 & [0.083, 0.111] & 0.043 & [0.030, 0.056] & 0.162 & [0.139, 0.185] & 0.267 & [0.240, 0.294]  \\
                9  & Park et al.~\cite{park2022_t6b}      & 0.075 & [0.063, 0.088] & 0.033 & [0.022, 0.044] & 0.127 & [0.106, 0.148] & 0.208 & [0.183, 0.233]  \\
                10 & Baseline                             & 0.061 & [0.049, 0.072] & 0.026 & [0.016, 0.036] & 0.102 & [0.083, 0.121] & 0.176 & [0.152, 0.200]  \\
                \bottomrule
            \end{tabular}
            \caption{Evaluation results with 95\% confidence intervals for the top system of each team.}
            \label{tab:evaluation_results}
        \end{table*}

        \section{Baseline System}\label{sec:baseline-system}

        In this section, we describe the task baseline system, as illustrated in Figure~\ref{fig:baseline_system}\@.
        The baseline system is a simplified version of the audio-text aligning framework presented in~\cite{Xie2022Unsupervised}, which calculates relevant scores between encoded textual descriptions (i.e., encoded captions) and encoded audio signals.
        It consists of two input encoders: one for audio, and the other for text, as illustrated in Figure~\ref{fig:baseline_system}.
        These two modality-specific encoders generate vector representations (i.e., audio embeddings and caption embeddings) for audio clips and textual descriptions.
        Then, the relevance score between an audio clip and a textual description is calculated by the dot product of their vector representations.
        The baseline system is optimized with a sampling-based triplet loss~\cite{Bromley1993Signature} at the training stage, and then applied to retrieve audio for caption queries at the testing stage.

        \subsection{Audio Encoder}\label{subsec:audio-encoder}

        A convolutional recurrent neural network (CRNN)~\cite{Xu2019ACRNN} is used as the audio encoder, which extracts frame-wise acoustic embeddings from audio signals.
        It consists of five convolution blocks, followed by a bidirectional gated recurrent unit (BiGRU).
        Each convolution block includes an initial batch normalization, a convolutional layer with padded $3 \times 3$ convolutions, and a LeakyReLU activation with a slope of $-0.1$.
        After the first, third, and fifth convolution blocks, one L4-Norm subsampling layer is used to reduce the temporal dimension of each block\textquotesingle s output by a factor of four.
        A dropout layer with a rate of $0.3$ is placed between the last L4-Norm layer and the BiGRU.
        Lastly, an up-sampling operation is applied to ensure the final output has the same temporal dimension as the CRNN input.

        The CRNN audio encoder takes 64-dimensional log mel-band energies as input.
        Each audio clip is split into 40 ms Hanning-windowed frames with a hop length of 20 ms.
        Then, 64 log mel-band coefficients are extracted from each frame.
        A sequence of 300-dimensional frame-wise acoustic embeddings are generated for each audio clip.
        The final audio embedding is calculated by averaging the frame-wise acoustic embeddings.

        \subsection{Text Encoder}\label{subsec:text-encoder}

        Word2Vec (Skip-gram model)~\cite{Mikolov2013Efficient} is utilized as the text encoder to convert textual descriptions into sequences of word embeddings.
        It is a two-layer fully-connected neural network, which learns word embeddings that are good at predicting surrounding words in a sentence or a document.
        For the sake of simplicity, we adopt a publicly available pre-trained Word2Vec~\cite{Word2Vec_online}, which is trained on Google News dataset.
        It consists of 300-dimensional word embeddings for roughly three million case-sensitive English words and phrases.
        The Word2Vec text encoder converts textual descriptions into sequences of semantic word embeddings word by word.
        The final caption embedding is computed by averaging the word embeddings.

        \subsection{Training Objective}\label{subsec:training-objective}

        The baseline system is trained by optimizing a ranking-based criterion~\cite{Bromley1993Signature}\@, such that audio clips and captions that belong together are more similar in the embedding space than mismatched audio-caption pairs.
        Specifically, across a batch of $N$ audio-caption pairs $\{(x_{n},y_{n})\}_{n=1}^{N}$, where $y_{n}$ is the caption pertaining to an audio clip $x_{n}$, we randomly select an imposter clip $\hat{x}_{n}$ and an imposter caption $\hat{y}_{n}$ for each audio-caption pair $(x_{n},y_{n})$.
        Then, the widely used sampling-based triplet loss~\cite{Xie2022Unsupervised, Harwath2016Unsupervised} is calculated by
        \begin{equation}
            \label{eq:triplet_loss}
            \begin{split}
                loss = \dfrac{1}{N} \sum_{n=1}^{N} & [ \max (0, S(x_{n},\hat{y}_{n}) - S(x_{n},y_{n}) + 1)      \\
                & + \max (0, S(\hat{x}_{n},y_{n}) - S(x_{n},y_{n}) + 1) ],
            \end{split}
        \end{equation}
        where $S$ is the audio-caption relevance score.

        \subsection{Baseline Results}\label{subsec:baseline-results}

        The baseline system is trained with batches of 32 audio-caption pairs in the training split for at most 150 epochs, while monitoring the loss~\eqref{eq:triplet_loss} on the validation split during the training process.
        An Adam optimizer with an initial learning rate of $0.001$ is adopted to optimize the training process.
        The learning rate is reduced by a factor of ten once the validation loss does not improve for five epochs.
        Training is terminated by early stopping with ten epochs.

        As shown in Table~\ref{tab:baseline_results}, the baseline system achieves similar performance in terms of mAP@10 and recall scores on the testing split and the evaluation dataset.
        Specifically, with the evaluation dataset, the theoretical chance levels are $1/{1000}=0.001$ for R@1, $1/{200}=0.005$ for R@5, and $1/{100}=0.01$ for R@10, respectively.
        In contrast to the theoretical chance levels, the baseline system obtains better recall scores, with an R@1 / R@5 / R@10 of $0.026$ / $0.102$ / $0.176$.
        The experimental results show that the baseline system can retrieve audio with their corresponding captions, i.e., perform language-based audio retrieval.
        On the other hand, since the baseline system employs a simple pipeline (e.g., averaging frame-wise acoustic embeddings and word embeddings), the retrieval performance remains limited.

        \begin{table}[!t]
            \centering
            \begin{tabular}{c||cccc}
                \toprule
                \bfseries Dataset & \bfseries mAP@10 & \bfseries R@1 & \bfseries R@5 & \bfseries R@10 \\
                \midrule
                Testing split     & 0.068            & 0.032         & 0.109         & 0.188          \\
                Evaluation        & 0.061            & 0.026         & 0.102         & 0.176          \\
                \bottomrule
            \end{tabular}
            \caption{Baseline results on the testing split and the evaluation dataset.}
            \label{tab:baseline_results}
        \end{table}

        \section{Challenge Submissions}\label{sec:challenge-submissions}

        In this section, we present the evaluation results and analysis of the submissions for language-based audio retrieval.

        \subsection{Evaluation Results}\label{subsec:evaluation-results}

        The task received a total number of 31 submissions from nine teams, with maximum four submissions per team allowed.
        Table~\ref{tab:evaluation_results} shows the results of evaluation metrics (e.g., mAP@10) with 95\% confidence intervals (CIs) for the top system of each team, comparing with the baseline system.
        The 95\% confidence intervals for evaluation metrics are calculated using the Jackknife estimate~\cite{Mesaros2019Sound}.

        All the submitted systems outperformed the baseline system in terms of mAP@10 and R@\textit{k} with \textit{k} $\in \{1, 5, 10\}$.
        Particularly, \mbox{Xu et al.}~\cite{xu2022_t6b} achieved the best performance, with a mAP@10 of $0.276$ and a 95\% CI between $0.254$ and $0.299$.
        \mbox{Mei et al.}~\cite{mei2022_t6b} ranked second, with their best system having a mAP@10 of $0.251$ (95\% CI $[0.229, 0.273]$).
        The top two teams obtained mAP@10 over $0.250$ and R@10 over $0.500$, in contrast to those of teams ranked third - fifth, having mAP@10 around $0.220$ and R@10 around $0.470$.
        Out of the nine teams, six teams achieved mAP@10 over $0.180$, R@5 over $0.300$, and R@10 over $0.410$.
        The baseline system ranked last with a significantly lower performance.

        \begin{table*}[!t]
            \centering
            \begin{tabular}{rr||p{0.25\linewidth}lp{0.23\linewidth}}
                \toprule
                \textbf{Rank} & \textbf{Team}                        & \textbf{Audio Modelling}                                                                                                                                                    & \textbf{Caption Modelling}                                & \textbf{Loss Function}                                    \\
                \midrule
                1             & Xu et al.~\cite{xu2022_t6b}          & PANNs~\cite{Kong2020PANNs}                                                                                                                                                  & BERT~\cite{Devlin2019BERT}, RoBERTa~\cite{Liu2019RoBERTa} & InfoNCE loss~\cite{Oord2018Representation}               \\
                2             & Mei et al.~\cite{mei2022_t6b}        & PANNs~\cite{Kong2020PANNs}                                                                                                                                                  & BERT~\cite{Devlin2019BERT}                                & NT-Xent loss~\cite{Chen2020ASimple}                      \\
                3             & Lamort et al.~\cite{lamort2022_t6b}  & PANNs~\cite{Kong2020PANNs}, OpenL3~\cite{Arandjelovic2017Look}, \newline VGGSound~\cite{Chen2020VGGsound}, VGGish~\cite{Hershey2017CNN}, \newline BART~\cite{Lewis2020BART} & Sentence-BERT~\cite{Reimers2019Sentence}                  & Triplet loss~\cite{Oncescu2021Audio, Xie2022Unsupervised}                     \\
                4             & Pellegrini~\cite{pellegrini2022_t6b} & PaSST~\cite{Koutini2021Efficient}                                                                                                                                           & Sentence-BERT~\cite{Reimers2019Sentence}                  & Triplet loss~\cite{Oncescu2021Audio, Xie2022Unsupervised}                  \\
                5             & Lai et al.~\cite{lai2022_t6b}        & ESResNet~\cite{Guzhov2020ESResNet}                                                                                                                                          & Transformer~\cite{Radford2021Learning}                                & CE loss~\cite{Radford2021Learning}                  \\
                6             & Wu et al.~\cite{wu2022_t6b}          & PANNs~\cite{Kong2020PANNs}, HTS-AT~\cite{Chen2022HTSAT}                                                                                                                     & Transformer~\cite{Radford2021Learning}                                & CE loss~\cite{Radford2021Learning}                            \\
                7             & Weck et al.~\cite{weck2022_t6b}      & PANNs~\cite{Kong2020PANNs}                                                                                                                                                  & RoBERTa~\cite{Liu2019RoBERTa}                             & Contrastive loss~\cite{Khosla2020Supervised}  \\
                8             & Xiao et al.~\cite{xiao2022_t6b}      & PANNs~\cite{Kong2020PANNs}                                                                                                                                                  & Word2Vec~\cite{Mikolov2013Efficient}                      & Triplet loss~\cite{Oncescu2021Audio, Xie2022Unsupervised}                  \\
                9             & Park et al.~\cite{park2022_t6b}      & CRNN~\cite{Xu2019ACRNN}                                                                                                                                                     & Word2Vec~\cite{Mikolov2013Efficient}                      & Triplet loss~\cite{Oncescu2021Audio, Xie2022Unsupervised}                 \\
                10            & Baseline                             & CRNN~\cite{Xu2019ACRNN}                                                                                                                                                     & Word2Vec~\cite{Mikolov2013Efficient}                      & Triplet loss~\cite{Oncescu2021Audio, Xie2022Unsupervised} \\
                \bottomrule
            \end{tabular}
            \caption{Summary of systems with best performance from all teams.}
            \label{tab:system_summary}
        \end{table*}

        \subsection{Analysis of Submissions}\label{subsec:analysis-of-submissions}

        The following analysis is based on the information reported by participating teams.

        \textbf{System summary}.
        All the 31 submitted systems and the baseline system adopted a bi-encoder architecture, which consisted of an audio encoder and a caption encoder, to associate audio with captions.
        For audio encoders, pre-trained convolutional neural networks (e.g., PANNs~\cite{Kong2020PANNs}) were the most common choice (27 systems from eight teams) among the nine participating teams.
        For caption encoders, pre-trained Transformer-based language embedding models (e.g., BERT~\cite{Devlin2019BERT} and Sentence-BERT~\cite{Reimers2019Sentence}) were frequently employed (28 systems from seven teams).
        Then, audio and captions were encoded into a common embedding space, where their relevance was scored with cosine similarity (24 systems from six teams) or dot product (seven systems from three teams) of their embeddings.
        Most of the submitted systems (23 systems from seven teams and the baseline) were trained by optimizing some contrastive losses (e.g., InfoNCE loss~\cite{Oord2018Representation} and triplet loss~\cite{Oncescu2021Audio, Xie2022Unsupervised}).

        \textbf{Audio encoders}.
        Among the submitted systems, PANNs~\cite{Kong2020PANNs} (e.g., CNN14, Wavegram-Logmel-CNN14 and CNN10) were the most common choice for encoding audio data.
        There were in total 21 systems from six teams utilizing PANNs as their audio encoders, including the top three teams~\cite{xu2022_t6b, mei2022_t6b, lamort2022_t6b}.
        Particularly, the CNN14 model was most preferred (20 systems).
        \mbox{Xu et al.}~\cite{xu2022_t6b} experimented with multiple pre-trained audio expert models, including CNN14 and Wavegram-Logmel-CNN14.
        With ensemble of different models, they achieved the best performance.
        \mbox{Mei et al.}~\cite{mei2022_t6b} adopted CNN14 as their audio encoder without model ensembles, having their best system ranked second.
        Tied on third rank, \mbox{Lamort et al.}~\cite{lamort2022_t6b} experimented with aggregating the most number of pre-trained audio expert models (i.e., five models in total, as shown in Table~\ref{tab:system_summary}).
        The audio encoders utilized in other systems were PaSST~\cite{Koutini2021Efficient}, ESResNet~\cite{Guzhov2020ESResNet}, HTS-AT~\cite{Chen2022HTSAT}, and CRNN~\cite{park2022_t6b}.

        Log-mel energies were frequently taken as input among the submitted systems.
        A total number of 28 systems from eight teams utilized log-mel energies as standalone audio features or in combination with other features (e.g., raw waveform~\cite{xu2022_t6b}).
        Other audio features used included log-magnitude spectrogram~\cite{lamort2022_t6b} and log-power spectrogram~\cite{lai2022_t6b}.

        \textbf{Caption encoders}.
        Most of the participating teams (i.e., seven out of nine) preferred Transformer-based language embedding models (e.g., BERT~\cite{Devlin2019BERT}) over word embedding models (e.g., Word2Vec~\cite{Mikolov2013Efficient}), to learn caption embeddings.
        As shown in Table~\ref{tab:system_summary}, pre-trained Transformer-based language embedding models, including BERT~\cite{Devlin2019BERT}, RoBERTa~\cite{Liu2019RoBERTa}, and Sentence-BERT~\cite{Reimers2019Sentence}, were utilized as caption encoders by seven teams (28 systems in total).
        The other two teams and the baseline converted captions into embeddings with pre-trained Word2Vec word vectors~\cite{Mikolov2013Efficient}.
        All teams with Transformer-based models ranked higher than those with pre-trained Word2Vec~\cite{Mikolov2013Efficient}, which showed that Transformer-based models learned caption embeddings more efficiently.

        \textbf{Loss functions}.
        The loss functions utilized in the submissions can be categorized into three groups: contrastive loss (one team), triplet loss (four teams), and N-pair loss (four teams).
        Contrastive loss~\cite{Sohn2016Improved} takes pairs of samples as input and measures the similarity between two inputs.
        Triplet loss~\cite{Oncescu2021Audio, Xie2022Unsupervised} takes as input triplets of samples, consisting of one anchor sample along with one positive and one negative sample.
        Compared to contrastive loss, triplet loss focuses on the difference of (dis-)similarities between positive and negative samples to the anchor sample.
        N-pair loss~\cite{Sohn2016Improved} shares a similar spirit to triplet loss, but extends to multiple negative samples, i.e., one positive and multiple negative samples for an anchor sample.
        In practice, several versions of N-pair loss were adopted, including InfoNCE loss~\cite{Oord2018Representation}, NT-Xent loss~\cite{Chen2020ASimple}, and symmetric cross-entropy (CE) loss~\cite{Radford2021Learning}.

        \textbf{Data augmentation and external data}.
        Data augmentation was adopted by only three participating teams (i.e., \mbox{Mei et al.}~\cite{mei2022_t6b} on rank two, \mbox{Lamort et al.}~\cite{lamort2022_t6b} on rank three, and \mbox{Wu et al.}~\cite{wu2022_t6b} on rank six) during their systems\textquotesingle~training.
        In contrast, external audio and textual data was leveraged for optimizing system performance by most participating teams (i.e., seven out of nine).
        The submitted systems were usually pre-trained with a large amount of external audio and textual data, and then fine-tuned on the task development dataset.
        For example, the top two teams (i.e., \mbox{Xu et al.}~\cite{xu2022_t6b} and \mbox{Mei et al.}~\cite{mei2022_t6b}) pre-trained their systems with AudioCaps~\cite{Kim2019AudioCaps}.
        Additionally, \mbox{Wu et al.}~\cite{wu2022_t6b} trained their systems by involving various external data, including audio samples and text (e.g., labels, file names and captions) from seven external datasets in addition to Clotho v2~\cite{Drossos2020Clotho}.

        \section{Conclusions}\label{sec:conclusions}

        The DCASE 2022 Challenge has introduced the language-based audio retrieval task, which is about using natural language textual captions as queries to retrieve audio signals from a dataset.
        This paper describes setups of the task, including task description, datasets for system development and evaluation, evaluation metrics, and a task baseline system.
        Moreover, this paper reports the final evaluation results of the task submissions and their analysis.
        The evaluation results show that all submitted systems outperformed the baseline in terms of evaluation metrics, with top performance being 0.276 in mAP@10 and 0.536 in R@10.

        One immediate observation about the submitted systems is that all of them adopted a bi-encoder architecture consisting of an audio encoder and a caption encoder.
        Pre-trained convolutional neural networks (e.g., PANNs~\cite{Kong2020PANNs}) were commonly employed as audio encoders, and pre-trained Transformer-based language embedding models (e.g., BERT~\cite{Devlin2019BERT}) were the preferred caption encoders.
        In addition to the task development dataset, external audio and textual data (e.g., labels and captions) was also frequently leveraged during system training.

        \bibliographystyle{IEEEtran}
        \bibliography{refs}
%
%
%
%
%
%
%
%
%

    \end{sloppy}
\end{document}